\documentclass[final,5p,times,twocolumn,authoryear]{elsarticle}

\usepackage{amssymb}
\usepackage{amsmath}
\usepackage{placeins}
\usepackage{booktabs}
\usepackage{adjustbox}
\usepackage{stfloats}
\usepackage{textgreek}

\journal{Astronomy \& Computing}

\begin{document}

\begin{frontmatter}

\title{Native Missing-Data Handling for Quasar Photometric Redshifts in Multiwavelength Surveys}

\author[first]{Mostafa Ahmed}
\author[first]{John Y. H. Soo\corref{cor1}}
\cortext[cor1]{Corresponding author.}
\ead{johnsooyh@usm.my}
\affiliation[first]{organization={School of Physics, Universiti Sains Malaysia},postcode={11800 USM},state={Pulau Pinang},country={Malaysia}}

\begin{abstract}
Photometric redshift estimation is essential for exploiting the large quasar samples produced by modern imaging surveys, but incomplete multi-band coverage can create severe colour--redshift degeneracies. We compare native missing-data handling with median, KNN, and flux-interpolation completion methods for full-scale spectroscopic samples from the Sloan Digital Sky Survey (SDSS) combined with samples from the Galaxy Evolution Explorer (GALEX), Wide-Field Infrared Survey Explorer (WISE), and the Panoramic Survey Telescope and Rapid Response System (Pan-STARRS1). The primary models, Masked Neural Networks (MNNs) and XGBoost, use raw photometric bands only and each handle missing data in a different way. Magnitude-dependent missing data are injected at 10\%, 20\%, and 30\%. For SDSS complete raw-band data, the performance of the MNN achieves $\sigma_{\rm RMS}=0.148$ and an outlier fraction of $13.7\%$, compared with $0.158$ and $16.9\%$ for XGBoost. For Pan-STARRS, the corresponding values are $0.177$ and $20.4\%$ for MNN, compared with $0.183$ and $22.7\%$ for XGBoost. At 30\% missing data, the MNN outlier fraction increases to $26.5\%$ for SDSS and $33.5\%$ for Pan-STARRS. Median imputation, the strongest completion baseline at 30\%, gives outlier fractions of $30.0\%$ and $35.2\%$, respectively. Handling missing data as a feature as a whole (XGBoost and MNN) therefore degrades gracefully and remains preferable to the tested completion baselines using imputation for its more reliable metrics and scatter, even at 30$\%$ missing data.
\end{abstract}


\end{frontmatter}

\section{Introduction}
\label{introduction}

\noindent Redshift is one of the central quantities used to map the Universe. It connects observed sources to distance, cosmic time, luminosity, and large-scale structure, therefore underpinning a wide range of extragalactic and cosmological applications. Spectroscopic redshifts provide the most reliable measurements, but spectroscopy is expensive in telescope time and is usually limited to brighter or specifically targeted objects. Photometric redshifts, or photo-$z$s, are therefore essential for exploiting the much larger statistical samples produced by modern imaging surveys \citep{salvato2019many}.

Quasars are a particularly important and difficult case. They trace luminous accretion onto supermassive black holes and can be observed across very large redshift ranges, making them valuable probes of the distant Universe \citep{combes2022active}. At the same time, their spectral energy distributions are not as straightforward as those of many normal galaxies. Galaxy photo-$z$ estimation often benefits from the prominent $4000\, \mathrm{\mathring{A}}$ break, while quasars show broad emission lines, absorption features, and a relatively smooth power-law continuum. These properties can produce severe colour-redshift degeneracies, where objects at different redshifts occupy similar regions of colour space \citep{salvato2019many}.

One practical way to reduce these degeneracies is to expand the wavelength coverage. Optical bands alone can struggle in certain redshift regimes, while ultraviolet and mid-infrared measurements can add information about different parts of the quasar spectral energy distribution. In this work, optical photometry from the Sloan Digital Sky Survey (SDSS) is combined with the $FUV/NUV$ bands from Galaxy Evolution Explorer (GALEX) and the $W1/W2$ measurements of the Wide-Field Infrared Survey Explorer (WISE), while the Panoramic Survey Telescope and Rapid Response System (Pan-STARRS) comparison sample uses $grizy$ photometry with WISE. This wider wavelength baseline is useful, but it creates an immediate complication: not every source has measurements in every band. Cross-matched catalogues inherit the footprint, depth, and detection limits of several surveys at once.

The "missing-ness" of data in this setting is not simply a random inconvenience. A missing band may correspond to a non-detection, a source falling outside a survey footprint, a bad or saturated measurement, or a band where the object is intrinsically too faint to be detected. This matters because the absence of a measurement can itself carry information about the source and the survey. A quasar that is missing in the ultraviolet, for example, is not necessarily equivalent to a quasar with an arbitrary table entry deleted. Treating all missing values as random can therefore erase part of the observational structure of the problem \citep{Rubin1976,Seaman2013}.

This work tests whether or not utilising missing features as a tool instead of treating it as random is a better approach. Related work has already used tree models that route missing values natively, mask-based neural networks for incomplete multivariate data, and machine-learning photo-$z$ methods for incomplete quasar photometry \citep{Chen2016,Che2018,Curran2022}. The contribution here is a controlled comparison of native handling and imputation for two full-scale quasar samples. This study asks the following questions: (1) how accurately can XGBoost and a masked neural network (MNN) estimate photo-$z$ values from raw incomplete bands; (2) how does the performance degrade as missing data increase from 10\% to 30\%; and (3) do median, $k$-nearest neighbour ($k$NN) or flux-interpolation baselines improve on native handling? These questions will be answered by the analysis on the various imputation methodologies reported below. 

The remainder of the paper is organised as follows: Section~2 describes the survey data and selection; Section~3 defines the missing-data experiments and models; Section~4 reports the results; Section~5 discusses interpretation, limitations, and applications; and Section~6 gives the conclusions and future work.

\section{Datasets}

\noindent This study uses two large, spectroscopically anchored quasar samples with different optical filter systems and auxiliary-survey coverage. SDSS provides the primary sample, while Pan-STARRS1 provides an independent comparison with a different footprint, depth and selection function. The final native-model test sets contain 96,285 SDSS objects and 22,868 Pan-STARRS objects. These counts refer to the held-out test samples; the parent catalogues are larger. The comparison is deliberately not an identical-band experiment: GALEX coverage is not uniform over the Pan-STARRS sample, so the Pan-STARRS baseline uses its native optical bands plus WISE rather than imposing a smaller common cross-match. The inputs for the SDSS quasar sample are $u,g,r,i,z,W1,W2,NUV,FUV$ while those for the Pan-STARRS quasar sample are $g,r,i,z,y,W1,W2$.

\subsection{Sloan Digital Sky Survey (SDSS)}

\noindent SDSS is a wide-field imaging and spectroscopic survey programme used for large-scale studies of galaxies, quasars, and the cosmic web. Data Release 19 (DR19) provides the parent catalogue for this work and contains approximately $750\,000$ spectroscopically confirmed quasars \citep{aghakhanloo_nineteenth_2026}. The raw optical inputs are the five SDSS bands $u,g,r,i,z$. The native-band runs add WISE $W1$ and $W2$, and retain the available GALEX bands described below. The final held-out test set contains 96,285 objects. Its concentration at low and intermediate redshift reflects flux limits, target selection, spectroscopic success, and quasar evolution rather than a volume-limited population.

Near-ultraviolet ($NUV$) and far-ultraviolet ($FUV$) measurements from GALEX are cross-matched to SDSS via right ascension and declination. In the prepared SDSS \texttt{TYPE=6} parent table, valid $NUV$ values are available for $221\,681$ of $481\,425$ objects ($46.05\%$) and valid $FUV$ values for $84\,649$ objects ($17.58\%$); WISE $W1$ and $W2$ values are available for more than $99.99\%$ of the same table. Since the ultraviolet coverage is incomplete, objects without $FUV$ or $NUV$ entries are retained with missing values rather than discarded. The SDSS sample therefore tests native handling under both real cross-survey incompleteness and controlled magnitude-dependent missing data. 

\subsection{Panoramic Survey Telescope and Rapid Response System (Pan-STARRS)}

\noindent Pan-STARRS1 (PS1) is a wide-field optical imaging survey with a large sky footprint and five broad bands. The value-added PS1--WISE quasar catalogue contains roughly $2.5$ million quasars with photometric redshifts; this work isolates its spectroscopically confirmed subset so that every evaluated object has a reference spec-$z$ \citep{KunsagiMate2022}. The final held-out test set contains $22\,868$ objects.

The raw PS1 feature space uses $g,r,i,z,y$, supplemented with WISE $W1$ and $W2$ bands \citep{Wright2010}. In the prepared PS1--WISE spectroscopic table, all $114\,338$ retained objects have valid $W1$ and $W2$ entries. This gives a useful comparison to SDSS while preserving the native PS1 optical system. GALEX is not imposed on this sample because uniform UV availability would reduce the usable sample and change the selection function; a direct GALEX-matched comparison is reserved for future work.

\subsection{Wide-field Infrared Survey Explorer (WISE)}

\noindent WISE provides the mid-infrared part of the feature space used in both survey samples. The mission observed the sky in four infrared bands, with the two shortest-wavelength channels, $W1$ and $W2$, centred at approximately $3.4$ and $4.6~\mathrm{\mu m}$ \citep{Wright2010}. These bands provide information beyond the optical and can help distinguish sources whose optical spectral energy distributions are degenerate. 

The same WISE information is used for both SDSS and Pan-STARRS, making it one of the common components in the inter-survey comparison. Its role is not to replace the optical bands, but to add a longer-wavelength anchor that helps separate objects whose optical colours alone can be ambiguous. This is particularly relevant for higher-redshift quasars, where emission features and continuum shape move through the observed optical filters.

\subsection{Galaxy Evolution Explorer (GALEX)}

\noindent GALEX provides the ultraviolet component of the SDSS feature set. The two GALEX bands used here are $FUV$ and $NUV$, which probe shorter wavelengths than the SDSS optical bands and are therefore useful for low-redshift colour degeneracies \citep{Morrissey2007}. In the early optical-only SDSS experiments, the $z\lesssim1$ redshift region showed a visible scatter tail in the $z_{\rm spec}$--$z_{\rm phot}$ plane. Adding the ultraviolet bands gave the models information from a part of the spectrum that was not represented by $ugriz$ alone, helping to reduce that degeneracy.

The GALEX component is also the clearest example of real cross-survey "missing-ness" in this work. Not every SDSS quasar has a corresponding $FUV$ or $NUV$ measurement, either because of footprint coverage, detection depth, source faintness, or catalogue quality cuts. These objects are not removed from the sample. Instead, the missing UV values are retained as \texttt{NaN}, allowing XGBoost and the MNN to work with the incomplete catalogue directly. This makes the SDSS + GALEX experiment a more realistic test case than a fully complete artificial table.

\begin{figure*}[!tbp]
\centering
\includegraphics[width=0.95\textwidth]{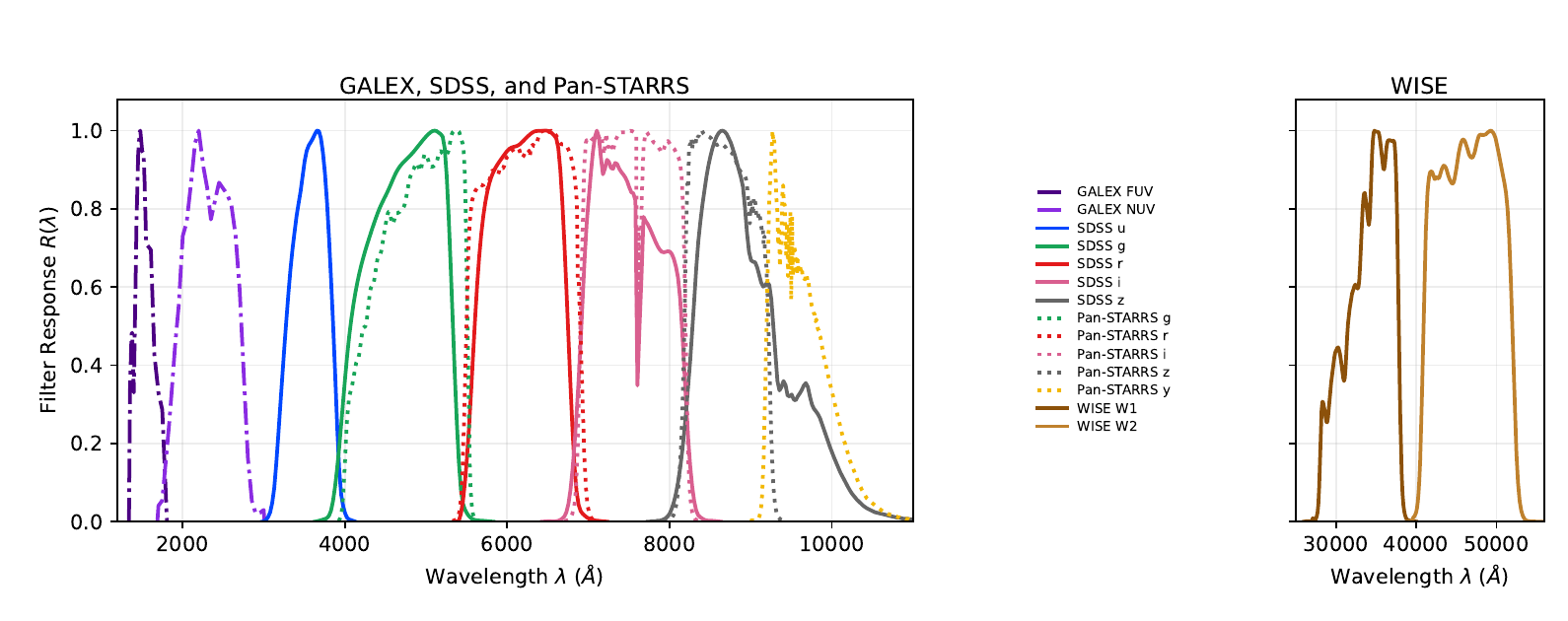}
\caption{Normalised response curves for the GALEX, SDSS, Pan-STARRS, and WISE filters, visualising the multi-wavelength feature coverage of quasars.}
\label{fig:filter_response}
\end{figure*}

The relative wavelength coverage of the survey filters mentioned above is shown in Figure~\ref{fig:filter_response}. The curves make the complementarity of the ultraviolet, optical, and mid-infrared inputs explicit; the model comparisons themselves use catalogue magnitudes rather than response curves as features.

\begin{figure*}[!tbp]
\centering
\includegraphics[width=0.95\textwidth]{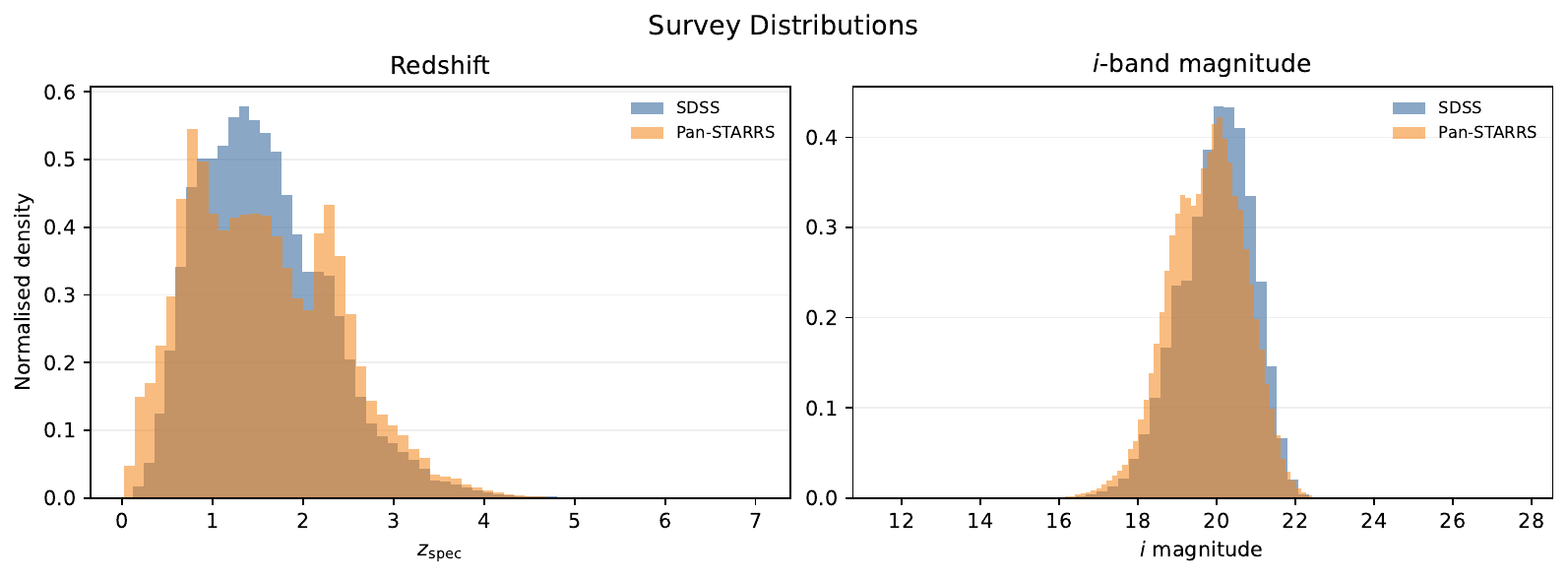}
\caption{Spectroscopic-redshift and common $i$-band magnitude distributions of the SDSS and Pan-STARRS samples.}
\label{fig:survey_distributions}
\end{figure*}

The final spectroscopic redshift and common-band magnitude distributions of the two survey samples are shown in Figure~\ref{fig:survey_distributions}. These distributions provide a direct view of the redshift coverage and optical magnitude range used for the subsequent training and testing experiments.

An important detail is that ``complete'' does not mean exactly the same thing for the two surveys. The Pan-STARRS complete case is effectively complete in the selected feature space. In contrast, the SDSS + GALEX catalogue already contains real missing $FUV$ and $NUV$ values because not every object is covered or successfully crossmatched by GALEX. Therefore, the SDSS complete case means that no additional synthetic missing data have been injected, not that every UV value is physically present.

The full results of photo-$z$ performance metrics for both the native missing-data handling models and imputation methods are shown in Appendix~\ref{app:full_scale_metrics}.

\begin{figure*}[!tbp]
\centering
\includegraphics[width=0.95\textwidth]{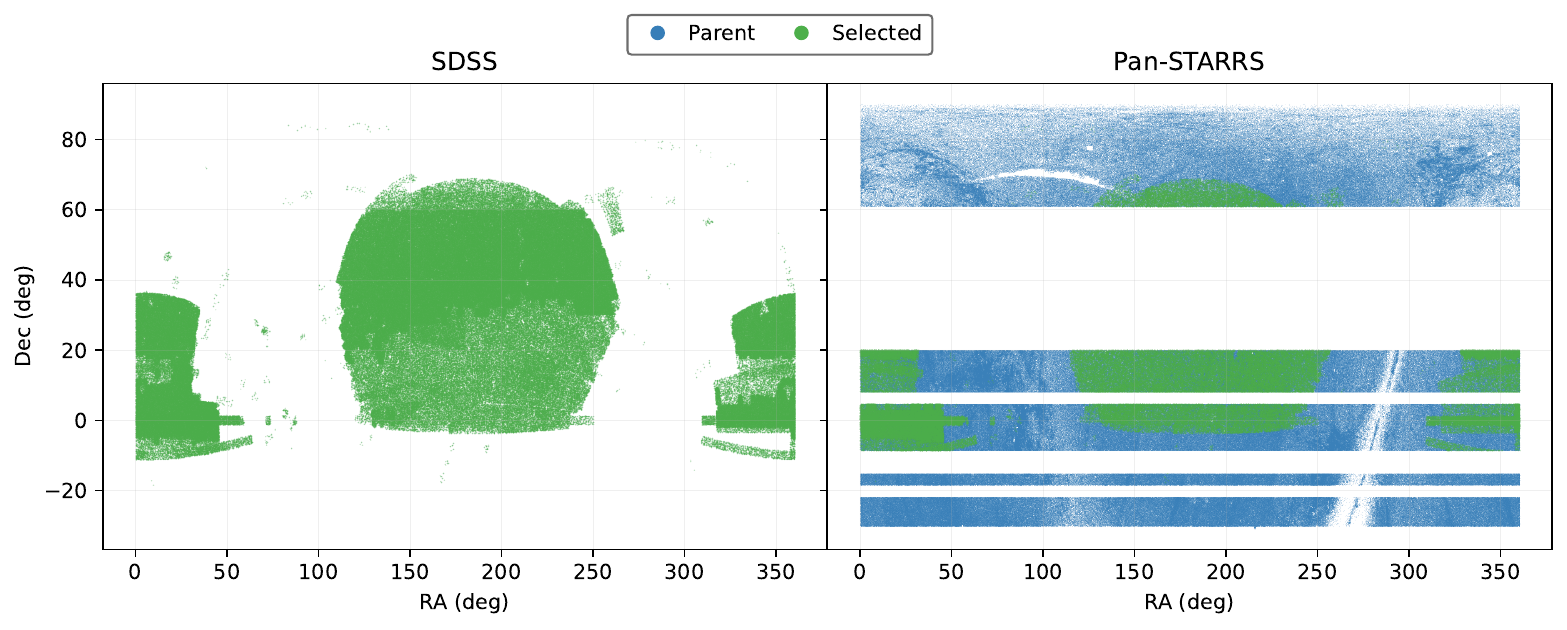}
\caption{Right ascension--declination coverage of the parent (blue) and selected (green) SDSS and Pan-STARRS samples.}
\label{fig:survey_radec}
\end{figure*}

\subsection{Sample Selection and Cross-Matching}

\noindent The analysis uses the prepared spectroscopic quasar tables supplied by the SDSS \texttt{TYPE=6} and PS1--WISE catalogue workflows. The SDSS parent table contains $481\,425$ \texttt{TYPE=6} rows and is restricted to $0<z_{\rm spec}<10$; the final held-out test set contains $96\,285$ objects. The PS1--WISE parent catalogue contains $2\,475\,775$ rows, of which $114\,338$ have usable spectroscopic redshifts in the same range; its held-out test set contains $22\,868$ objects. Catalogue sentinel values are converted to missing data rather than interpreted as measurements. The SDSS sample retains its available GALEX $FUV/NUV$ values, while PS1 retains WISE $W1/W2$. Right ascension and declination are retained for the parent and selected samples, allowing the footprint and cross-survey coverage to be checked directly. For the controlled SDSS feature comparisons, the object identities and held-out split remain fixed while the available band set is changed; differences between those runs therefore reflect feature coverage rather than a change in the evaluated objects. Footprint, depth, morphology, and spectroscopic-selection effects can affect both the missing-data pattern and the measured photo-$z$ performance.

The parent and selected-sample sky coverage is shown in Figure~\ref{fig:survey_radec}. This is a footprint and cross-survey coverage sanity check, not a proof that rows are identical between every catalogue. For the controlled SDSS feature comparisons, a separate identity audit confirms valid RA/DEC values, identical held-out spectroscopic-redshift ordering, and the same $96\,285$ evaluated rows across all feature sets. The selected samples inherit the non-uniform footprints of the contributing surveys, which is one reason "missing-ness" cannot be assumed to be independent of the observed catalogue properties.

\section{Methodology}

\noindent The testing procedure targets the missing-photometry aspect of quasar photo-$z$ estimation rather than the entire photo-$z$ problem. The spec-$z$s serve as reference values. The photometric magnitude bands are first cleaned; sentinel values are then represented as missing data; controlled magnitude-dependent missing data are then injected into the sample, creating three samples with $10\%$, $20\%$ and $30\%$ missing data, respectively. Native models predict from the incomplete inputs, while imputation baselines first estimate the absent values and then pass the completed table to XGBoost. All methods are compared using the same 80-20 training-testing split, plotting conventions, and performance metrics defined below.

\subsection{Missing Data}

\noindent Missing photometric measurements in astronomical catalogues are not usually equivalent to randomly deleted table entries. In the classical missing-data taxonomy introduced by \citet{Rubin1976}, data are \textit{missing completely at random} when the probability of missing data is unrelated to either observed or unobserved quantities, \textit{missing at random} when the "missing-ness" can be explained by observed quantities, and \textit{missing not at random} (MNAR) when the probability of absence is itself related to the missing value or to latent physical quantities associated with it. The latter case is particularly relevant for wide-field astronomical surveys: a source may lack a measurement in a given band not because of arbitrary catalogue failure, but because it lies outside a survey footprint, falls below a detection threshold, has low signal-to-noise in that wavelength range, or has a spectral energy distribution that makes it intrinsically faint in that band. For quasars, this is especially important because observed colours and band detections are coupled to redshift, emission-line placement, absorption features, dust, survey depth, and crossmatch availability.

In practice, missing values can result from non-detections, bad measurements, saturation, masked pixels, blending, footprint gaps, and failed crossmatches. These mechanisms differ physically, but all remove one or more usable inputs. The catalogue encodings used by the source tables, including sentinel values such as $-1$, $-99$,  $-9999$ or invalid-value flags, are converted to \texttt{NaN} before modelling. The models therefore treat the resulting missing entries uniformly, even though their observational origins are not identical.

Operationally, sentinel values in the catalogues are first converted to \texttt{NaN}, so that the "missing-ness" is represented explicitly, rather than by artificial numerical placeholders. For the controlled missing-data experiments, additional missing values are injected using a magnitude-dependent MNAR prescription. For each band, the masking probability is increased for fainter objects: in the scripts, the probability is scaled from the difference between the object magnitude and the bright-end reference percentile for that band, then normalised to reach the target missing data level of $10\%$, $20\%$, or $30\%$. Since larger magnitudes correspond to lower fluxes, this approximates the observational fact that faint measurements are more likely to be absent. Despite being a simple model, it is still more realistic than uniform random deletion because it ties "missing-ness" to detectability.

\subsection{Algorithms}

\noindent In this work, we compare the models using the same performance metrics and visual diagnostics to assess the reliability of the photo-$z$ estimates produced. The same framework may also generalise to datasets with similar photometric coverage. The following discusses the three main methods used in dealing with the missing data in the quasar samples.

\subsubsection{Imputation}

\noindent The imputation methods simply replace missing \texttt{NaN} values before the XGBoost is trained. Three methods are considered: flux interpolation, median imputation and $k$NN imputation. The aim is not to exhaust the large range of available imputers, but to test whether representative completion methods can compete with models that handle missing data directly.

The first baseline approach to missing fluxes is linear interpolation, which leverages the approximate continuity of a quasar's spectral energy distribution (SED). By interpolating between adjacent available photometric bands, one can approximate the missing flux. While computationally inexpensive and physically intuitive for smooth continua, this deterministic method can struggle to capture complex emission-line features or abrupt spectral breaks characteristic of quasars \citep{salvato2019many,Ivezic2014}. In this work, interpolation is performed in flux space after converting magnitudes to fluxes, and the interpolated values are converted back to magnitudes before training.

Next, median imputation was used as the simplest completion-based baseline. After replacing catalogue sentinel values with \texttt{NaN}, each missing value in a given photometric band was replaced by the median of the available values in that same band,
\[
x_{ij}^{\rm imp} =
\begin{cases}
x_{ij}, & \text{if } x_{ij} \text{ is observed},\\
{\rm median}\left(x_j\right), & \text{if } x_{ij} \text{ is missing},
\end{cases}
\]
where $x_{ij}$ is the magnitude of object $i$ in band $j$. This produces a fully complete table that can be passed to models or software that do not accept missing values. In the context of this project, median imputation does not attempt to reconstruct the object-specific spectral shape, nor does it use neighbouring bands to estimate the missing measurement. It simply assigns a typical value for that band to all objects with missing entries. This makes it useful as a conservative baseline. However, median imputation is limited because it reduces the variance of the affected bands and can weaken the colour relationships between filters, which is a known drawback of simple single-imputation methods \citep{LittleRubin2019,vanBuuren2018}.

Finally, $k$-nearest neighbours ($k$NN) imputation was deployed using $k=5$. This method estimates missing values from nearby training objects after scaling the available magnitudes \citep{Troyanskaya2001}. This is a fixed local baseline, not a claim that $k=5$ is globally optimal: it was chosen to limit averaging across distinct quasar populations while avoiding a single-neighbour estimate that is overly sensitive to noise. The completed $k=3,5,10$ sensitivity runs are retained with the supplementary results and are used to quantify the practical effect of this choice.

\subsubsection{XGBoost}

\noindent While the imputation techniques explicitly estimate and replace missing flux values prior to analysis, extreme gradient boosting (XGBoost) bypasses explicit imputation entirely \citep{Chen2016}. Rather than generating proxy values for missing photometric bands, XGBoost can perform regression directly on datasets with missing values. The core architecture of XGBoost relies on a tree ensemble model, which uses a set of classification and regression trees. This framework allows the algorithm to integrate the presence of missing data directly into its learning process, shifting the analytical focus away from repairing the dataset and toward optimising the predictive task in the presence of incomplete information \citep{An2018}.

The mechanism by which XGBoost natively handles missing data is built into its optimal split-finding algorithm during tree construction. When evaluating a specific feature at a given node, the algorithm does not require a substituted numerical value to proceed. Instead, XGBoost calculates an optimal default direction for instances lacking that feature, a pathway that is learned from the data. It evaluates the structure score of possible splits, including the impact of sending missing instances to the left branch versus the right branch, and selects the direction that maximises the gain \citep{Chen2016}. In the context of multi-wavelength quasar surveys, where missing fluxes often result from sources falling below detection limits or lacking coverage in specific bands, this means that the algorithm can learn whether the missing data itself carries predictive value \citep{ye_machine_2024}.

\subsubsection{Masked Neural Network (MNN)}

\noindent The MNN was used as a supervised neural-network model designed to predict $z_{\rm phot}$ from incomplete multi-band photometry. The main difference between this model and a standard neural network is that missing values are not treated as ordinary numerical inputs. Instead, each object is represented by both its photometric measurements and a binary mask that records which bands are available. Sentinel values are first converted to \texttt{NaN}; the observed magnitudes are then scaled, while missing entries are set to zero only after their absence has been recorded by the mask. This allows the network to distinguish between an actual measured value and a band that is unavailable.

This distinction is important in survey-based quasar photo-$z$ work, because missing photometry is often related to survey depth, sky coverage, wavelength coverage, or the intrinsic brightness and colour of the source, rather than being purely random. The MNN therefore learns from two sources of information: the measured magnitudes themselves and the pattern of missing bands. Rather than discarding incomplete objects or forcing them to take imputed values, the model keeps those objects in the training and testing samples and uses the mask to account for the incomplete feature space. Mask-based neural-network approaches are commonly used for learning from incomplete data \citep{Che2018}, and incomplete photometry is a known issue in quasar photometric-redshift estimation \citep{Curran2022}. During training, additional bands are randomly hidden from the model to improve robustness, following the same general idea as dropout regularisation \citep{Srivastava2014}. In this work, the MNN is therefore intended to test whether a photo-$z$ model can remain reliable when realistic survey photometry is partially missing, without relying on a separate imputation step.

\subsection{Model Configuration}

\noindent The experiments use an $80/20$ stratified train/test split with random seed $42$. XGBoost uses $1\,000$ estimators with maximum depth $6$, learning rate $0.05$, subsample and column-sample fractions of $0.8$, minimum child weight $3$, and the squared-error regression objective. The MNN uses hidden widths $256$--$128$--$64$, batch normalisation, leaky ReLU activation with slope $0.05$, dropout rate $0.3$, random feature masking probability of $0.2$, Adam optimiser with learning rate $10^{-3}$, batch size $512$, maximum $150$ epochs, and early stopping after $15$ epochs without validation improvement. The input to the MNN is the scaled photometric bands concatenated with a binary availability mask. Median, $k$NN ($k=5$), and flux interpolation are fitted only on the training split before XGBoost is trained on the completed table. The $k$NN sensitivity is evaluated at $k=3,5,10$. The model configurations are summarised in Table~\ref{tab:model_settings}. Generative AI tools (Codex, most recently GPT-6-Luna, OpenAI; Claude, most recently Opus 5.5, Anthropic) assisted with code development, data-analysis troubleshooting, and figure preparation. All code and results were reviewed and verified by the authors.

\begin{table}[!t]
\centering
\caption{Principal model settings used in this work.}
\label{tab:model_settings}
\begin{tabular}{lll}
\toprule
Model & Parameters & Values \\
\midrule
XGBoost & Estimators / depth & 1000 / 6 \\
XGBoost & Learning rate & 0.05 \\
XGBoost & Subsample / columns & 0.8 / 0.8 \\
MNN & Hidden layers & 256, 128, 64 \\
MNN & Dropout / mask probability & 0.3 / 0.2 \\
MNN & Optimiser / learning rate & Adam / $10^{-3}$ \\
Both & Random seed & 42 \\
\bottomrule
\end{tabular}
\end{table}

\subsection{Performance Metrics}

\noindent Comparative plots of $z_\mathrm{phot}$ vs. $z_\mathrm{spec}$ are complemented by three performance metrics. Together, they measure the typical residual size, the robust scatter, and the fraction of catastrophic failures.

\subsubsection*{Root-Mean-Square (RMS) Error}

\noindent The RMS error ($\sigma_{\rm RMS}$) is a standard deviation-like metric that measures the typical size of photometric redshift errors by averaging the squared residuals between $z_{\rm spec}$ and $z_{\rm phot}$ and then taking the square root \citep{Ilbert2006}. In photo-$z$ work it is often defined as
\begin{equation}
    \sigma_{\rm RMS} = \sqrt{\left\langle \left( \frac{z_{\rm phot} - z_{\rm spec}}{1 + z_{\rm spec}} \right)^2 \right\rangle}.
\end{equation}
Since it uses squared residuals and a mean, $\sigma_{\rm RMS}$ is sensitive to catastrophic outliers, and therefore tends to be larger when even a small fraction of very poor photo-$z$ estimates is present.

\subsubsection*{Normalised Median Absolute Deviation (NMAD)}

NMAD ($\sigma_{\rm NMAD}$) is a robust scatter metric commonly used to assess photometric redshift accuracy, defined as
\begin{equation}
    \sigma_{\rm NMAD} = 1.4826 \times \text{median}\left(\frac{|\Delta z - \text{median}(\Delta z)|}{1 + z_{\rm spec}}\right),
\end{equation}
where $\Delta z = z_{\rm phot} - z_{\rm spec}$ \citep{Brammer2008}. By using the median instead of the mean, $\sigma_{\rm NMAD}$ minimises the influence of catastrophic outliers that would otherwise inflate RMS-based error estimates.

\subsubsection{Outlier Fraction}

The outlier fraction used in this work is the standard catastrophic outlier fraction commonly denoted by $\eta$ in photometric-redshift studies \citep{Ilbert2006,Brinchmann2017}. For each object, the normalised redshift residual is
\begin{equation}
    \Delta z_i = \frac{z_{{\rm phot},i}-z_{{\rm spec},i}}{1+z_{{\rm spec},i}} .
\end{equation}
An object is counted as a catastrophic outlier if
\begin{equation}
|\Delta z_i| > 0.15 .
\end{equation}
The outlier fraction is then calculated as
\begin{equation}
\eta = \frac{N\left(|\Delta z|>0.15\right)}{N_{\rm total}}\times 100 ,
\end{equation}
where $N\left(|\Delta z|>0.15\right)$ is the number of objects whose normalised residual exceeds the threshold and $N_{\rm total}$ is the total number of tested objects. These are therefore the $\eta$ outliers in the usual photo-$z$ sense: they are not based on a Gaussian $\sigma$ cut, but on a fixed threshold in normalised redshift error.

\section{Results}\label{sec:res}

\noindent The full results of photo-$z$ performance metrics for both the native missing-data handling models and imputation methods are shown in Appendix~\ref{app:full_scale_metrics}. The full-scale experiments were carried out on the SDSS DR19 \texttt{TYPE=6} + GALEX/WISE sample and the Pan-STARRS1 + WISE spectroscopic sample. The SDSS test set contains $96\,285$ objects, while the Pan-STARRS test set contains $22\,868$ objects. The main aim is to test whether native XGBoost and MNN remain stable under increasing levels of magnitude-dependent missing data, and whether imputation improves on native handling.

\subsection{Native Missing-Data Models}

\begin{figure*}[!tbp]
\centering
\includegraphics[width=0.95\textwidth]{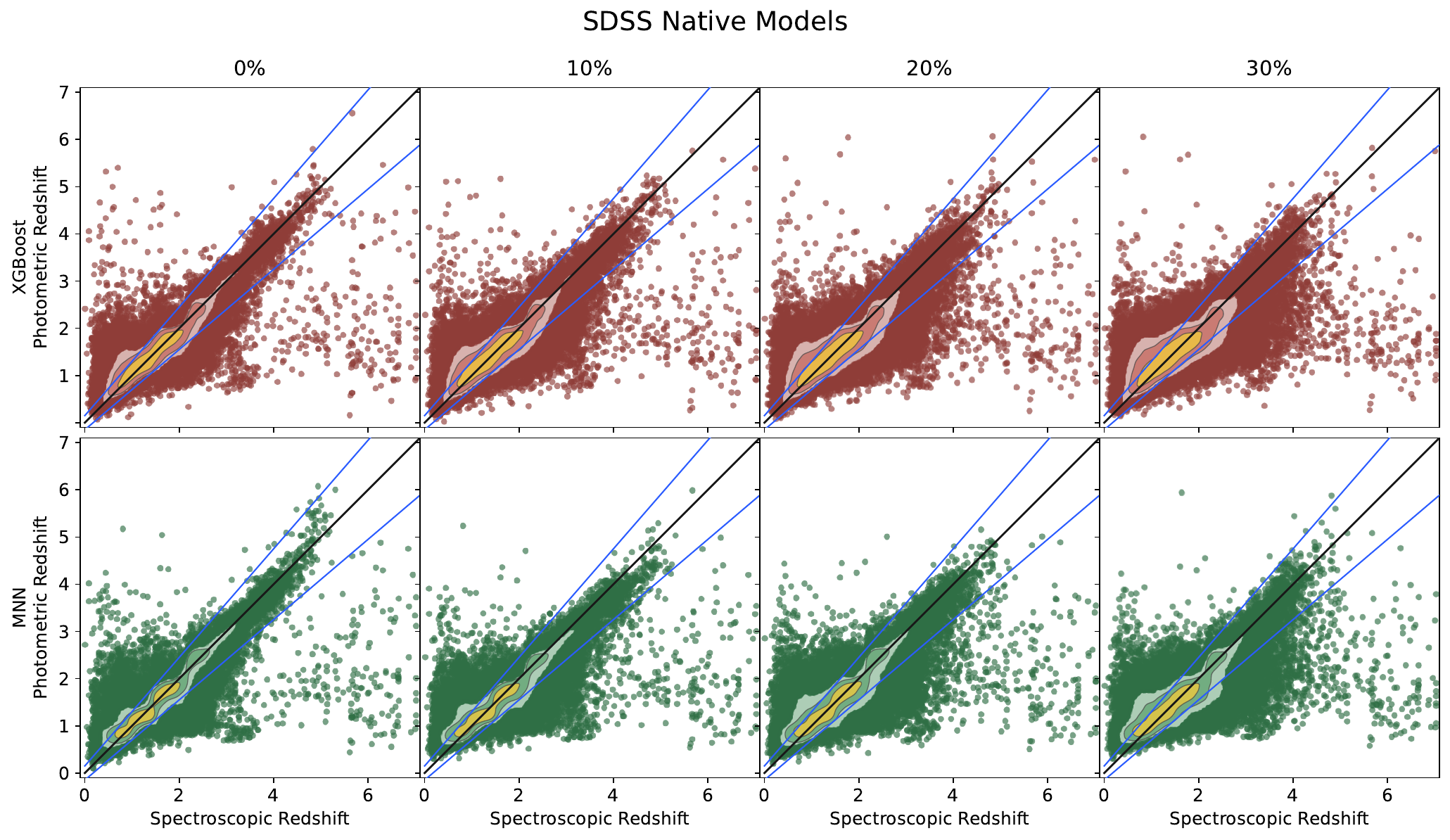}
\caption{$z_\mathrm{phot}$ vs. $z_\mathrm{spec}$ of the SDSS quasar sample. Columns from left to right show $0\%$,  $10\%$, $20\%$ and $30\%$ injected missing data, respectively; while the rows from top to bottom show the results of XGBoost and MNN.}
\label{fig:sdss_native_grid}
\end{figure*}

\begin{figure*}[!tbp]
\centering
\includegraphics[width=0.95\textwidth]{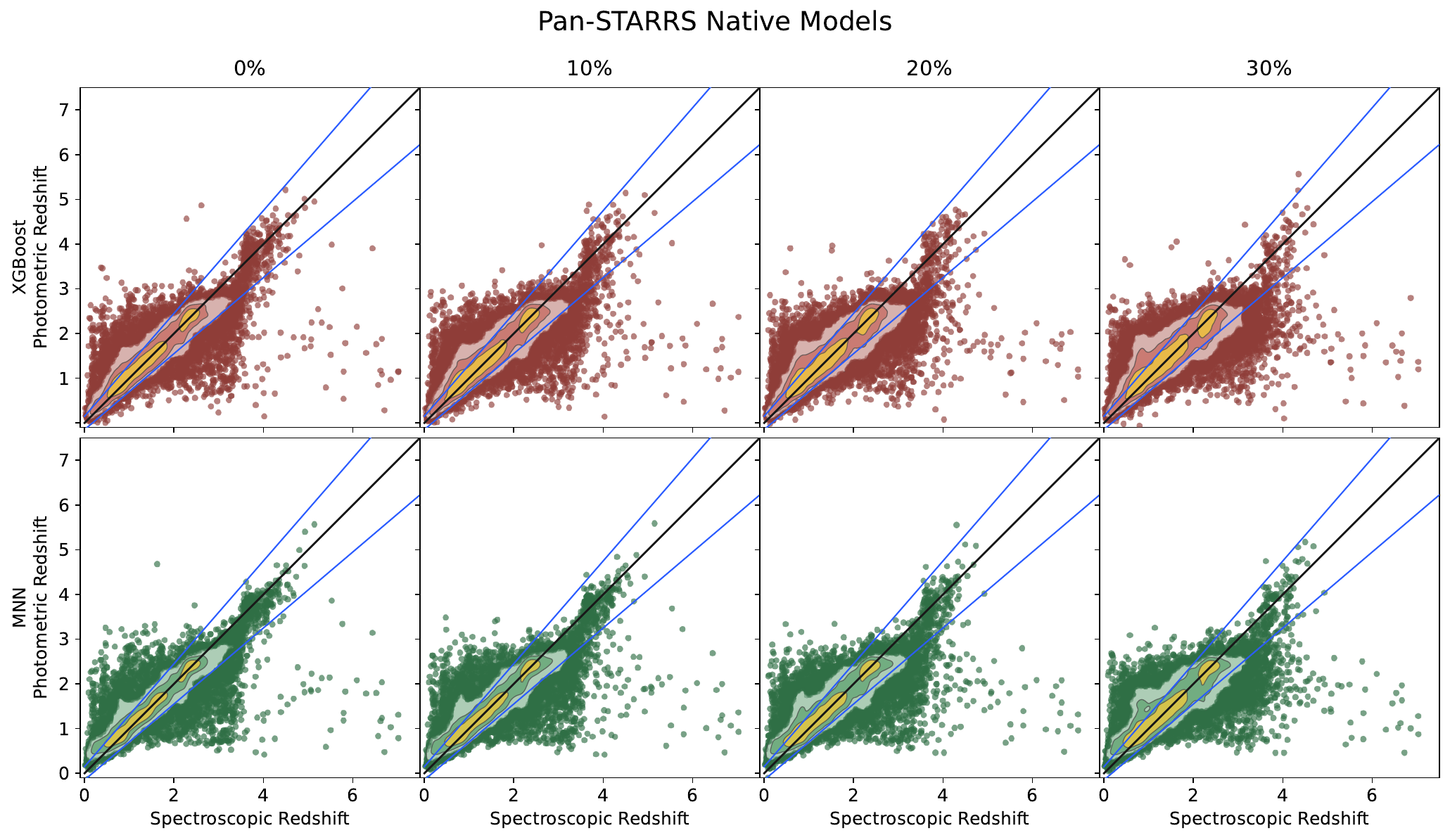}
\caption{$z_\mathrm{phot}$ vs. $z_\mathrm{spec}$ of the Pan-STARRS quasar sample. Columns from left to right show $0\%$,  $10\%$, $20\%$ and $30\%$ injected missing data, respectively; while the rows from top to bottom show the results of XGBoost and MNN.}
\label{fig:panstarrs_native_grid}
\end{figure*}

\noindent For the native missing-data models, the results are shown in Figures~\ref{fig:sdss_native_grid} and~\ref{fig:panstarrs_native_grid}. In the SDSS \texttt{TYPE=6} + GALEX/WISE sample, both XGBoost and the MNN produce a clear one-to-one structure between $z_{\rm spec}$ and $z_{\rm phot}$. The MNN gives $\sigma_{\rm RMS}=0.148$, NMAD $=0.071$, and an outlier fraction of $13.7\%$, compared to $0.158$, $0.085$, and $16.9\%$ for XGBoost. At $30\%$ missing data, the MNN reaches $\sigma_{\rm RMS}=0.189$ and an outlier fraction of $26.5\%$, while XGBoost reaches $0.199$ and $28.8\%$. The degradation is measurable but gradual rather than catastrophic.

The Pan-STARRS1 + WISE results show the same broad trend, although the absolute metrics are slightly worse than SDSS. The MNN gives $\sigma_{\rm RMS}=0.177$, NMAD $=0.086$, and an outlier fraction of $20.4\%$, compared to $0.183$, $0.103$, and $22.7\%$ for XGBoost. At $30\%$ missing data, the MNN reaches $(\sigma_\mathrm{RMS},\sigma_\mathrm{NMAD},\eta)=(0.221, 0.141, 33.5\%)$, while XGBoost reaches $(0.226,0.149,34.2\%)$. The MNN advantage is therefore strongest in SDSS and more modest in Pan-STARRS, which is consistent with the different band coverage and sample selection.

\subsection{Imputation Models}

\begin{figure*}[!tbp]
\centering
\includegraphics[width=0.95\textwidth]{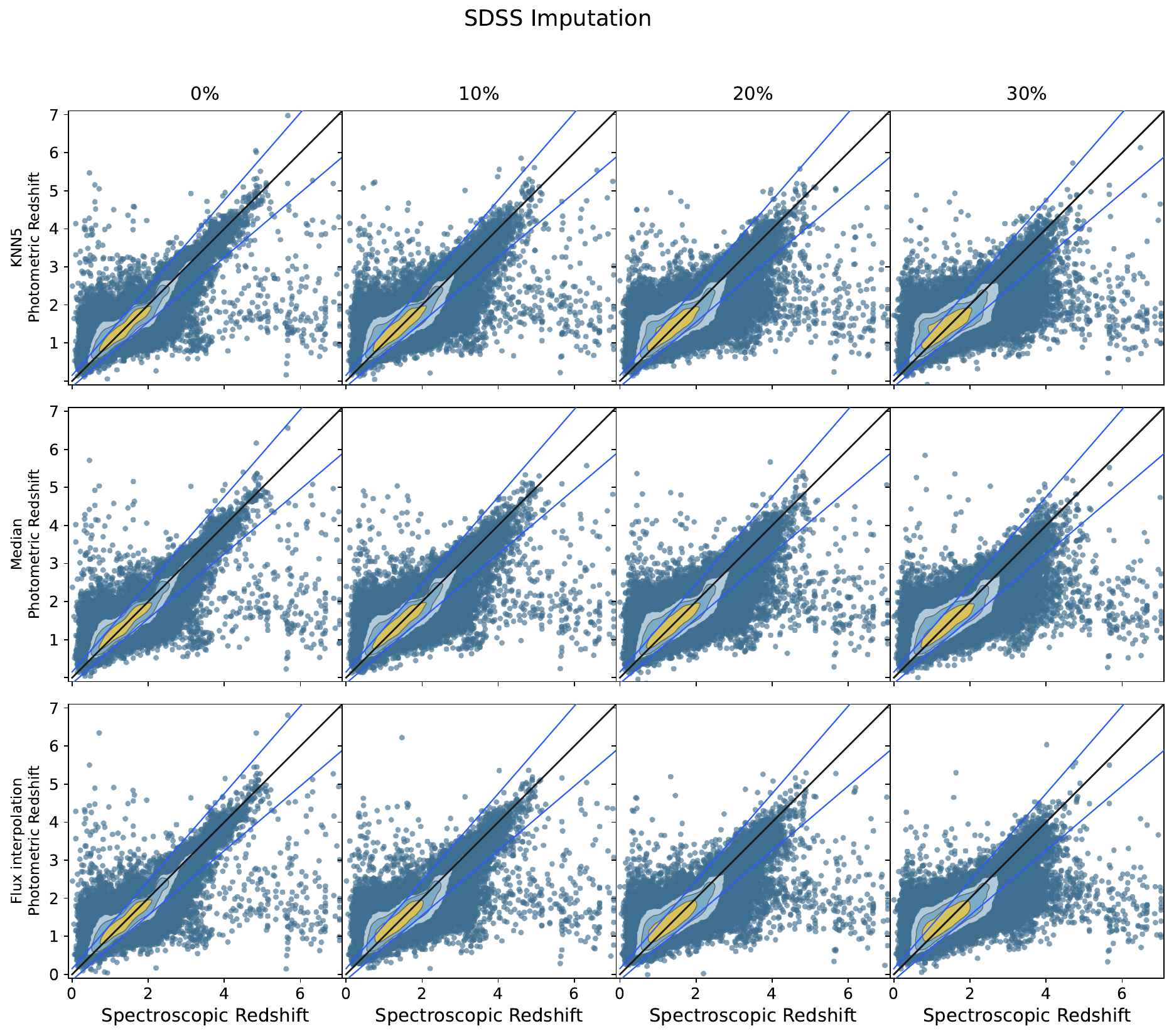}
\caption{$z_\mathrm{phot}$ vs. $z_\mathrm{spec}$ of the SDSS quasar sample using three imputation methods: $k$NN (top), median (middle) and flux interpolation (bottom). Columns from left to right show $0\%$,  $10\%$, $20\%$ and $30\%$ injected missing data, respectively.}
\label{fig:sdss_raw_imputation}
\end{figure*}

\begin{figure*}[!tbp]
\centering
\includegraphics[width=0.95\textwidth]{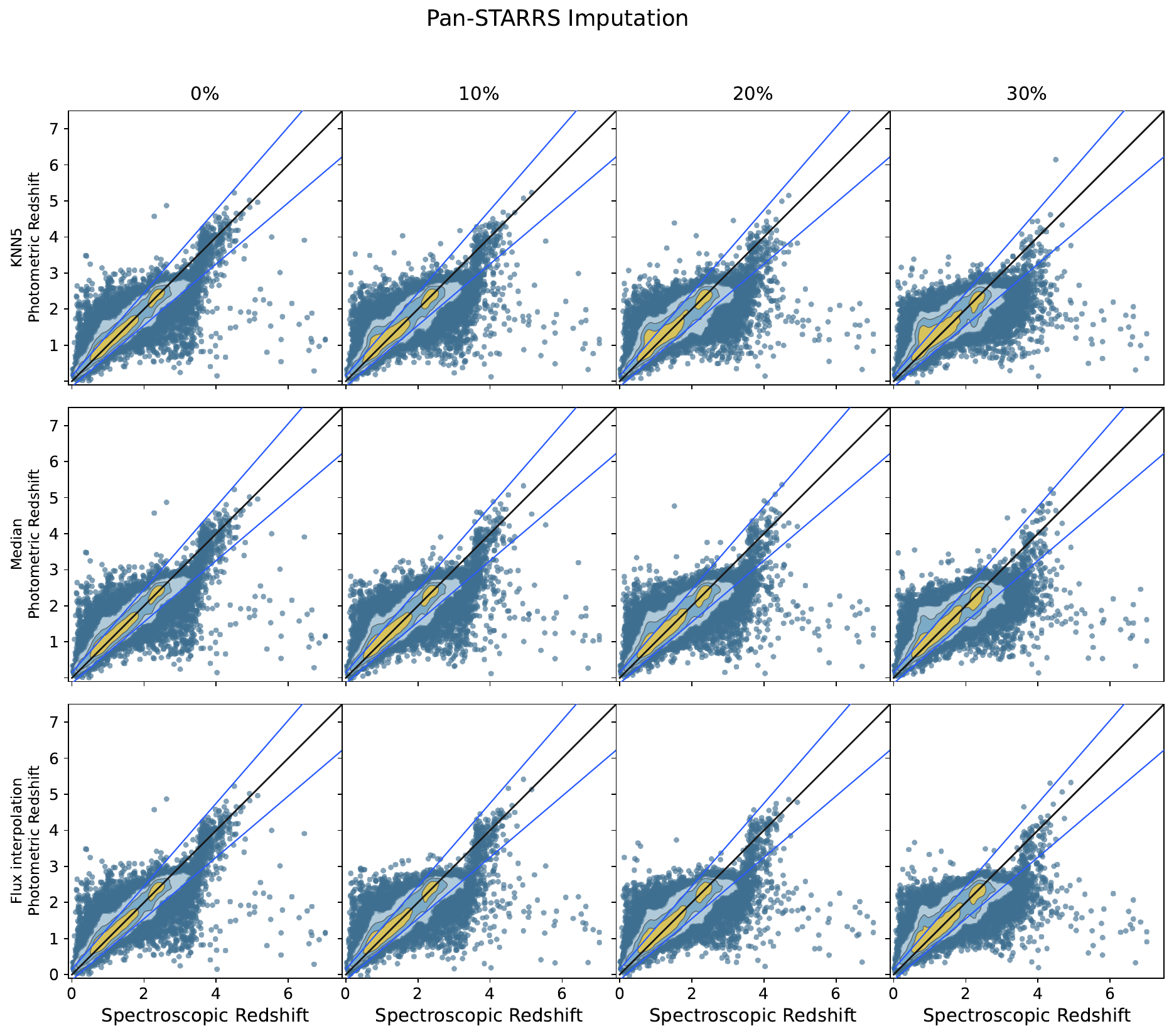}
\caption{$z_\mathrm{phot}$ vs. $z_\mathrm{spec}$ of the Pan-STARRS quasar sample using three imputation methods: $k$NN (top), median (middle) and flux interpolation (bottom). Columns from left to right show $0\%$,  $10\%$, $20\%$ and $30\%$ injected missing data, respectively.}
\label{fig:panstarrs_raw_imputation}
\end{figure*}

\noindent The photo-$z$ results for imputation methods are shown in Figures~\ref{fig:sdss_raw_imputation} and~\ref{fig:panstarrs_raw_imputation}. Median, $k$NN with $k=5$, and flux interpolation are all followed by XGBoost. Median imputation is the strongest of the three baselines at $30\%$ in both surveys, but it remains worse than native handling: in SDSS it gives $\sigma_{\rm RMS}=0.202$ and $\eta=30.0\%$, compared to $0.199$ and $28.8\%$ for native XGBoost; in Pan-STARRS it gives $0.229$ and $35.2\%$, compared to $0.226$ and $34.2\%$. The imputation methods therefore do not improve the native missing data handling models. 

We also conducted a $k$NN sensitivity test, which provides a quantitative check on the choice of $k$. Averaged over the 10\%, 20\%, and 30\% cases, the SDSS values of $(\sigma_{\rm RMS},\sigma_{\rm NMAD},\eta)$ are $(0.1988,0.1269,28.96\%)$, $(0.1981,0.1269,28.78\%)$, and $(0.1972,0.1256,28.48\%)$ for $k=3$, $5$, and $10$, respectively. The corresponding Pan-STARRS values are $(0.2195,0.1424,32.97\%)$, $(0.2193,0.1422,32.87\%)$, and $(0.2191,0.1418,32.83\%)$. Relative to $k=3$, increasing to $k=10$ changes the average RMS by only $0.82\%$ in SDSS and $0.18\%$ in Pan-STARRS, and changes the outlier fraction by only $1.67\%$ and $0.42\%$, respectively. Thus $k=10$ is marginally best in this limited test, while $k=5$ is already within about one percent of both alternatives. We retain $k=5$ as a representative moderate-neighbourhood baseline rather than treating it as an optimised value: lower values offer no measurable advantage here, and increasing the neighbourhood produces only small gains.

\section{Discussion and Potential Applications}

\subsection{The "Missing-ness" of Data as Predictive Information}

\noindent The main result of this work is not simply to identify which model performs better than the other; the broader point here is that missing photometry in quasar surveys should not always be treated as a nuisance that must be erased before modelling. In real surveys, a missing band often has a physical or observational reason behind it: the source may be too faint in that wavelength range, outside the footprint of a survey, below the detection threshold, or affected by the way different catalogues overlap. This makes the missing data partly informative, which is why the MNAR framework is more appropriate than treating the missing values as random table defects \citep{Rubin1976}. In this work, XGBoost and the MNN benefit from keeping this structure available to the model. XGBoost does this through learned default directions in its trees \citep{Chen2016}, while MNN does it more explicitly through binary masks that tell the network which bands are present and which are not, following the general idea that missing data patterns can themselves carry predictive information \citep{Che2018}.

\subsection{Role of Multi-wavelength Feature Coverage}

\noindent The SDSS results in Section~\ref{sec:res} show how strongly the feature space controls the quality of the photo-$z$ solution. In previous works, optical-only attempts produced a visible low-redshift scatter tail, especially around $z\lesssim1$. This is the region where different quasar colours can become degenerate when only the optical bands are available. Adding GALEX $FUV$ and $NUV$ measurements helped reduce this behaviour by giving the models direct access to the ultraviolet part of the spectral energy distribution \citep{Morrissey2007}. The improvement was not restricted to a small visual change; it also improved the overall stability of the predictions. Similarly, the inclusion of WISE $W1$ and $W2$ magnitudes extended the wavelength baseline into the mid-infrared, which is particularly useful when optical colours alone are not enough to separate redshift regimes \citep{Wright2010,salvato2019many}. In other words, the algorithms are important, but the bands are not secondary. The model cannot recover wavelength information that was never represented in the input features.

\begin{figure*}[!tbp]
\centering
\includegraphics[width=0.95\textwidth]{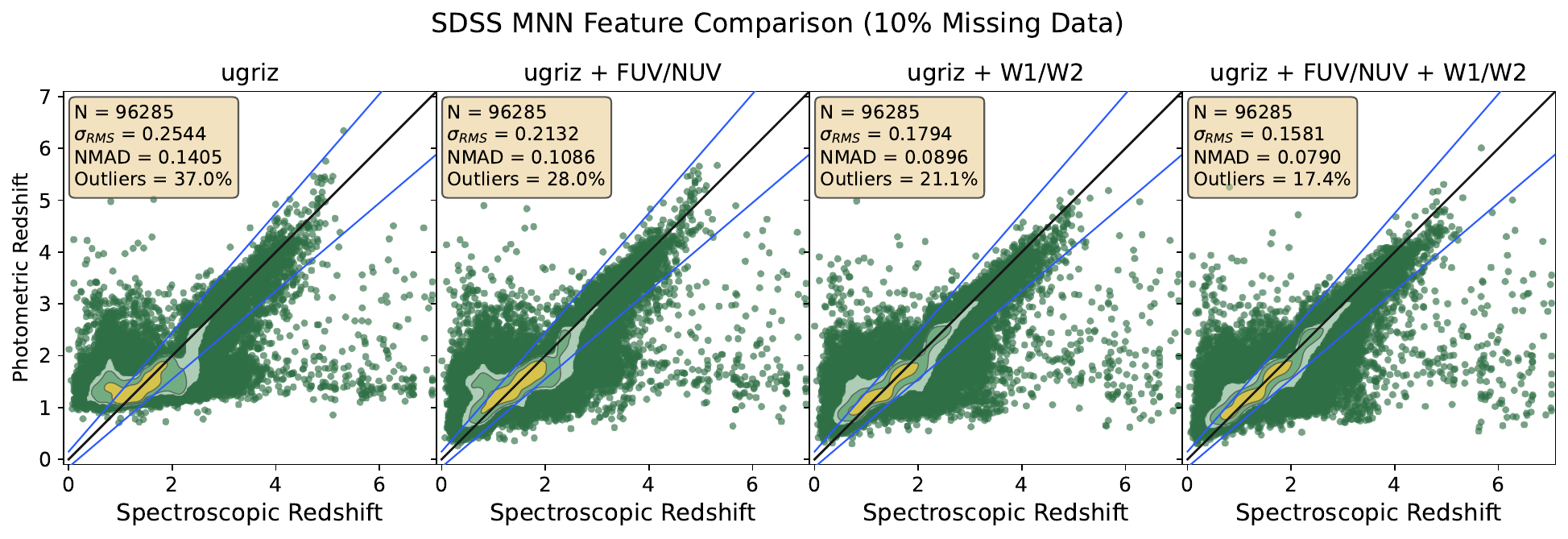}
\caption{Photo-$z$ vs. spec-$z$ on a controlled SDSS feature-inclusion comparison on the same held-out objects using the masked neural network (MNN), with $10\%$ injected missing data. The four panels show results using the magnitude bands $ugriz$, $ugriz+FUV/NUV$, $ugriz+W1/W2$, and $ugriz+FUV/NUV+W1/W2$, respectively.}
\label{fig:sdss_mnn_feature_ablation}
\end{figure*}

As a controlled feature-coverage diagnostic, we hold the MNN fixed while varying the SDSS input band magnitudes on the same held-out test split, using the sample with $10\%$ injected missing data; the corresponding photo-$z$ scatter comparison is shown in Figure~\ref{fig:sdss_mnn_feature_ablation}. All four runs keep the random feature-masking augmentation used in the other missing-data experiments, so they differ only in their input bands. The MNN gives $\sigma_{\rm RMS}=0.254$, $\sigma_{\rm NMAD}=0.140$, and an outlier fraction of $37.0\%$ using $ugriz$ alone, improving to $0.213$, $0.109$, and $28.0\%$ with $FUV/NUV$, $0.179$, $0.090$, and $21.1\%$ with $W1/W2$, and $0.158$, $0.079$, and $17.4\%$ when all band magnitudes are included. Relative to $ugriz$ alone, the full band set therefore reduces $\sigma_{\rm RMS}$ by $38\%$, $\sigma_{\rm NMAD}$ by $44\%$, and the outlier fraction by $53\%$, even though measurements are missing throughout the sample. This reduction of $19.6$ percentage points in the outlier fraction is about five times larger than the $3.7$-point increase caused by injecting $10\%$ missing data into the full band set, from $13.7\%$ to $17.4\%$. The full band set at $30\%$ missing data ($26.5\%$) also still outperforms $ugriz$ at $10\%$. Wider wavelength coverage therefore matters more here than the loss of $10\%$ of the measurements. Because the model, source table, and evaluation split are held fixed, this diagnostic assesses feature coverage rather than a change in the evaluated objects or model family. The $ugriz$ panel shows a sharp lower boundary near $z_{\rm phot}\approx1$. This is not imposed on the predictions, which are not clipped; it reflects regression toward the conditional mean under the optical colour--redshift degeneracy, sharpened by the masking augmentation. Disabling the augmentation for this configuration lowers the boundary, with $2\,832$ rather than $73$ predictions below $z_{\rm phot}=1$, while leaving the outlier fraction almost unchanged at $37.2\%$.

\subsection{Current UV Coverage}

\noindent The SDSS+GALEX case is also where the limitations of the current data become clearest. A large number of objects still have missing $FUV$ and $NUV$ values because GALEX does not cover the entire SDSS quasar sample. The fact that the MNN and XGBoost still perform well under these conditions is encouraging, but it also suggests that the results could improve further if wider and deeper UV coverage were available. Future UV facilities such as ULTRASAT are relevant in this context because they are designed to provide wide-field near-ultraviolet coverage at much larger survey grasp than GALEX \citep{shvartzvald_ultrasat_2024}. ULTRASAT is not a direct replacement for the full GALEX $FUV$+$NUV$ combination, but broader $NUV$ coverage would still reduce one of the major missing-data gaps in this type of crossmatched quasar catalogue.

\subsection{Imputation Compared with Native Handling}

\noindent The comparison with imputation remains an important part of the study. Median imputation is cheap and stable, KNN is more adaptive but more computationally demanding, and flux interpolation is physically motivated but can force a value into a band where the absence of a detection is itself informative. From our results, median is the strongest completion baseline at $30\%$, but native XGBoost and the MNN remain better in both surveys. KNN with $k=5$ is weaker than median in both surveys, and generally weaker than flux interpolation except at low missing-data levels in SDSS. The $k=3,5,10$ sensitivity test shows that the exact neighbourhood size has only a small effect here; $k=10$ is slightly better on average, while $k=5$ remains a reasonable fixed baseline because it preserves a relatively local neighbourhood without relying on a single neighbour.

\subsection{Computational Considerations}

\noindent From a practical point of view, XGBoost and the MNN occupy different roles. XGBoost is easier to run, quick to set up, and robust with very little additional machinery. This makes it useful as a strong baseline and as a practical model for large catalogue work. On the other hand, MNN requires more care: the magnitudes must be scaled, the masks must be built correctly, and the training loop is more involved. However, the reward is visible in the scatter plots and metrics: the MNN gives lower $\sigma_{\rm RMS}$, NMAD, and outlier fractions than XGBoost at every missing-data level in both surveys. In this sense, XGBoost is the simpler production-ready option, while the MNN is the stronger missing-data model when the extra setup is acceptable.

This trade-off has received little direct attention in the photo-$z$ literature, where missing values are typically imputed \citep[e.g.][]{Curran2022} or handled by training separate models for different band combinations \citep[e.g.][]{Tian2026}. Our results are nevertheless consistent with related work. Native missing-value support in gradient-boosted trees outperformed imputation at lower computational cost in a benchmark of health databases \citep{PerezLebel2022}, and a neural-network regression improved on XGBoost for galaxy photo-$z$s from the DESI Legacy Imaging Surveys \citep{Tian2026}. In contrast, tree-based models remained state of the art across general tabular benchmarks of about $10\,000$ samples \citep{Grinsztajn2022}, far smaller than the training splits used here, although we did not test the effect of sample size directly.

\subsection{Potential Applications}

\noindent The possible applications go beyond simply generating photo-$z$ values. A model that can handle incomplete photometry allows astronomers to keep sources that would otherwise be discarded because one or more bands are missing. This is useful for constructing larger quasar catalogues, prioritising spectroscopic follow-up, identifying high-redshift candidates, estimating redshift distributions in incomplete sky regions, and preparing target lists for future surveys. It could also be useful in cross-survey studies where no single catalogue provides uniform coverage across the full sky. The key advantage is not that the method replaces spectroscopy; spectroscopic redshifts remain the reference standard. Rather, spectroscopically confirmed quasars are a small fraction of those detected in imaging surveys. In the PS1--WISE parent catalogue used here, only $114\,338$ of $2\,475\,775$ sources ($4.6\%$) have a usable spectroscopic redshift \citep{KunsagiMate2022}, and SDSS imaging alone yields over a million photometrically selected quasar candidates, of which fewer than $90\,000$ had spectroscopic confirmation \citep{Richards2009}. Samples of this kind already support science that the spectroscopic subsets cannot provide on their own: the clustering of about $800\,000$ SDSS photometric quasars has been used to constrain primordial non-Gaussianity \citep{Leistedt2014}, and the all-sky Quaia catalogue provides 1.3 million quasars with redshifts estimated from low-resolution Gaia spectra and unWISE photometry for cosmological analyses \citep{StoreyFisher2024}. Photometric redshifts also determine which candidates receive spectroscopic time, as in searches for rare $z>5.6$ quasars, where photometric preselection precedes spectroscopic confirmation \citep{Banados2016}. A method that keeps sources with missing bands in such samples, rather than discarding them, therefore enlarges the population available for these analyses.

\subsection{Limitations}

\noindent Despite the potential applications in photo-$z$ studies and beyond, this work has its limitations. The models are trained and evaluated only where spectroscopic redshifts exist, so the validation set inherits the selection effects of the spectroscopic sample. The MNAR injection used here is also a simplified model of survey "missing-ness": it mimics the fact that fainter measurements are more likely to be absent, but real catalogues also contain footprint effects, depth variations, blending, source morphology issues, and crossmatch failures. Finally, the SDSS+GALEX results are limited by the available ultraviolet coverage. These limitations do not weaken the main result, but they define its scope: the method is best understood as a robust way to work with realistic incomplete photometric catalogues, not as a replacement for better data.

\section{Conclusion and Future Work}

This work tested whether quasar photometric redshifts can be estimated reliably when realistic missing photometry is present. Across data samples of SDSS DR19 + GALEX/WISE and Pan-STARRS1 + WISE, native missing-data models have been shown to produce promising results. In the complete case, MNN produces photo-$z$s with uncertainties as low as $\sigma_{\rm RMS}=0.148$ and $\eta=13.7\%$ for SDSS, compared to $0.158$ and $16.9\%$ using XGBoost. For Pan-STARRS, the corresponding values are $0.177$ and $20.4\%$ for MNN, compared to $0.183$ and $22.7\%$ for XGBoost.

The same behaviour remains visible under stronger missing data. At 30\%, the MNN gives $\eta=26.5\%$ in SDSS and $33.5\%$ in Pan-STARRS, while XGBoost gives $28.8\%$ and $34.2\%$, respectively. Thus, MNN reduces the outlier fraction by about $8.0\%$ relative to XGBoost in SDSS and $2.0\%$ in Pan-STARRS at the highest tested missing-data level. Median imputation, the strongest completion baseline at this level, gives outlier rates of $30.0\%$ and $35.2\%$, so it does not improve on native handling.

The results also show that feature coverage is central to the problem. GALEX $FUV/NUV$ helped reduce the low-redshift degeneracy seen in optical-only experiments, while WISE $W1$ and $W2$ extended the wavelength baseline into regimes where optical bands alone are less separable. At the same time, the SDSS+GALEX catalogue still contains many real UV non-detections or non-crossmatches, meaning that the models were tested under genuinely incomplete survey conditions. The main conclusion is therefore that incomplete photometry does not have to be treated only as a reason to discard objects or force imputation. When missing data is represented properly, especially through masking in MNN or native missing-value handling in XGBoost, the model can still recover useful photo-$z$ structure and keep partially observed quasars scientifically usable.

Several extensions would make the test more complete. The first is to repeat the same missing-data framework on additional large quasar catalogues with different selection functions and photometric systems. This would help separate behaviour that is genuinely model-driven from behaviour that is specific to SDSS or Pan-STARRS. A second step is to apply trained models to photometric-only quasar candidates, where no spectroscopic redshift is available, and then validate the most interesting subsets through external catalogues or follow-up spectroscopy. This would test the practical catalogue-building value of the method beyond the supervised training/testing settings used here.

A more detailed treatment of missing data is also needed. The MNAR prescription used in this work is intentionally simple and magnitude-dependent, but real surveys contain spatially varying depth, footprint gaps, detector artefacts, blending, saturation, and crossmatch failures. Future work should therefore test missing data models that include sky position, survey depth maps, quality flags, and band-specific detection limits. Wider ultraviolet coverage would also be valuable. The SDSS+GALEX results suggest that UV information is important for reducing the low-redshift scatter tail, but the present catalogue is limited by incomplete GALEX coverage. Future UV surveys, including facilities such as ULTRASAT, could provide a stronger test of how much quasar photo-$z$ performance improves when the UV part of the feature space becomes less sparse \citep{shvartzvald_ultrasat_2024}.

\section*{Acknowledgements}

Mostafa Ahmed and John Y. H. Soo acknowledge financial support via the Fundamental Research Grant Scheme (FRGS) by the Malaysian Ministry of Higher Education with code FRGS/1/2023/STG07/USM/02/14. Mostafa also acknowledges the numerous times he was constantly seeking the counsel of close friends and family on general matters that may help him with regards to writing this paper. In addition, we acknowledge the SDSS, Pan-STARRS1, GALEX, and WISE teams for making the survey data products publicly available. 

SDSS is managed by the Astrophysical Research Consortium for the Participating Institutions of the SDSS Collaboration, including the Carnegie Institution for Science, Chilean National Time Allocation Committee (CNTAC) ratified researchers, Caltech, the Gotham Participation Group, Harvard University, Heidelberg University, The Flatiron Institute,  The Johns Hopkins University, L’Ecole polytechnique fédérale de Lausanne (EPFL), Leibniz-Institut für Astrophysik Potsdam (AIP), Max-Planck-Institut für Astronomie (MPIA Heidelberg), Max-Planck-Institut für Extraterrestrische Physik (MPE), Nanjing University, National Astronomical Observatories of China (NAOC), New Mexico State University, The Ohio State University, Pennsylvania State University, Smithsonian Astrophysical Observatory, Space Telescope Science Institute (STScI), the Stellar Astrophysics Participation Group, Universidad Nacional Autónoma de México (UNAM), University of Arizona, University of Colorado Boulder, University of Illinois at Urbana-Champaign, University of Toronto, University of Utah, University of Virginia, Yale University, and Yunnan University. 

The Pan-STARRS1 Surveys (PS1) and the PS1 public science archive have been made possible through contributions by the Institute for Astronomy, the University of Hawaii, the Pan-STARRS Project Office, the Max-Planck Society and its participating institutes, the Max Planck Institute for Astronomy, Heidelberg and the Max Planck Institute for Extraterrestrial Physics, Garching, The Johns Hopkins University, Durham University, the University of Edinburgh, the Queen's University Belfast, the Harvard-Smithsonian Center for Astrophysics, the Las Cumbres Observatory Global Telescope Network Incorporated, the National Central University of Taiwan, the Space Telescope Science Institute, the National Aeronautics and Space Administration under Grant No. NNX08AR22G issued through the Planetary Science Division of the NASA Science Mission Directorate, the National Science Foundation Grant No. AST–1238877, the University of Maryland, Eotvos Lorand University (ELTE), the Los Alamos National Laboratory, and the Gordon and Betty Moore Foundation.

This publication makes use of data products from the Wide-field Infrared Survey Explorer, which is a joint project of the University of California, Los Angeles, and the Jet Propulsion Laboratory/California Institute of Technology, funded by the National Aeronautics and Space Administration.

Based on observations made with the NASA Galaxy Evolution Explorer. GALEX is operated for NASA by the California Institute of Technology under NASA contract NAS5-98034.

\section*{CRediT authorship contribution statement}
Mostafa Ahmed: Conceptualization, Methodology, Software, Formal analysis, Investigation, Visualization, Writing -- original draft. John Y. H. Soo: Supervision, Methodology, Writing -- review and editing, Funding acquisition.

\section*{Declaration of competing interest}
The authors declare that they have no known competing financial interests or personal relationships that could have appeared to influence the work reported in this paper.

\section*{Data availability}
The source survey catalogues are publicly available through the SDSS, Pan-STARRS1, GALEX, and WISE archive services cited in this paper. The derived prediction tables, metric tables, and analysis scripts generated for this study are available from the corresponding author upon reasonable request.

%\clearpage

\appendix

%%% APPENDIX A %%%
\section{Quasar Photometric Redshift Performance Metrics}\label{app:full_scale_metrics}
Tables~\ref{tab:sdss_metrics} and \ref{tab:panstarrs_metrics} show the full results for each algorithm and performance metrics, with $10\%$, $20\%$ and $30\%$ missing data.
\begin{table}[!tbp]
\centering
\caption{Photo-$z$ performance metrics for the SDSS quasar sample, at various percentages of missing data.}\label{tab:sdss_metrics}
\begin{tabular}{lrccc}
\toprule
Method & Missing & $\sigma_{\rm RMS}$ & $\sigma_{\rm NMAD}$ & $\eta$ (\%) \\
\midrule
MNN & 0\% & 0.148 & 0.071 & 13.7 \\
XGBoost & 0\% & 0.158 & 0.085 & 16.9 \\
KNN5 & 0\% & 0.163 & 0.089 & 18.0 \\
Median & 0\% & 0.159 & 0.086 & 17.3 \\
Flux interp. & 0\% & 0.164 & 0.089 & 18.3 \\
\midrule
MNN & 10\% & 0.158 & 0.079 & 17.4 \\
XGBoost & 10\% & 0.173 & 0.100 & 21.0 \\
KNN5 & 10\% & 0.184 & 0.109 & 24.1 \\
Median & 10\% & 0.176 & 0.104 & 22.1 \\
Flux interp. & 10\% & 0.182 & 0.110 & 24.2 \\
\midrule
MNN & 20\% & 0.177 & 0.095 & 21.3 \\
XGBoost & 20\% & 0.186 & 0.113 & 24.8 \\
KNN5 & 20\% & 0.198 & 0.127 & 28.9 \\
Median & 20\% & 0.189 & 0.118 & 26.0 \\
Flux interp. & 20\% & 0.196 & 0.127 & 28.6 \\
\midrule
MNN & 30\% & 0.189 & 0.113 & 26.5 \\
XGBoost & 30\% & 0.199 & 0.128 & 28.8 \\
KNN5 & 30\% & 0.213 & 0.145 & 33.3 \\
Median & 30\% & 0.202 & 0.133 & 30.0 \\
Flux interp. & 30\% & 0.207 & 0.142 & 32.5 \\
\bottomrule
\end{tabular}
% \begin{adjustbox}{width=\textwidth}
% \setlength{\tabcolsep}{4pt}
% \renewcommand{\arraystretch}{0.88}
% \input{Tables/raw_band_sdss_metrics.tex}
% \end{adjustbox}
\end{table}

\begin{table}[!tbp]
\centering
\caption{Photo-$z$ performance metrics for the Pan-STARRS quasar sample, at various percentages of missing data.}\label{tab:panstarrs_metrics}
\begin{tabular}{lrccc}
\toprule
Method & Missing & $\sigma_{\rm RMS}$ & $\sigma_\mathrm{NMAD}$ & $\eta$ (\%) \\
\midrule
MNN & 0\% & 0.177 & 0.086 & 20.4 \\
XGBoost & 0\% & 0.183 & 0.103 & 22.7 \\
\midrule
MNN & 10\% & 0.192 & 0.103 & 24.3 \\
XGBoost & 10\% & 0.198 & 0.120 & 26.9 \\
KNN5 & 10\% & 0.207 & 0.127 & 29.3 \\
Median & 10\% & 0.200 & 0.122 & 27.8 \\
Flux interp. & 10\% & 0.205 & 0.126 & 28.8 \\
\midrule
MNN & 20\% & 0.208 & 0.124 & 29.4 \\
XGBoost & 20\% & 0.211 & 0.133 & 30.3 \\
KNN5 & 20\% & 0.223 & 0.148 & 34.3 \\
Median & 20\% & 0.215 & 0.137 & 31.6 \\
Flux interp. & 20\% & 0.220 & 0.142 & 33.0 \\
\midrule
MNN & 30\% & 0.221 & 0.141 & 33.5 \\
XGBoost & 30\% & 0.226 & 0.149 & 34.2 \\
KNN5 & 30\% & 0.240 & 0.168 & 38.9 \\
Median & 30\% & 0.229 & 0.154 & 35.2 \\
Flux interp. & 30\% & 0.234 & 0.157 & 36.7 \\
\bottomrule
\end{tabular}
\end{table}

%\clearpage

\section*{Declaration of generative AI and AI-assisted technologies in the manuscript preparation process}
During the preparation of this work, the authors used Codex (OpenAI) and Claude (Anthropic) in order to assist with language editing and text organisation. After using these tools/services, the authors reviewed and edited the content as needed and take full responsibility for the content of the published article.

\bibliographystyle{elsarticle-harv}
\bibliography{references}

\end{document}